LETTER

# A 10-bit SAR ADC with 1.5× Input Range

Yi Zhang  [1, a]

**Abstract** This paper presents a differential 10-bit 2 MS/s successive approximation register (SAR) analog-to-digital converter (ADC) with a precision-improvement technique. The proposed method breaks the direct tradeoff between the capacitive digital-to-analog converter (CDAC) units and the resolution of SAR ADCs by extending the effective input range, thereby enhancing the effective number of bits (ENOB). Specifically, the technique is implemented by switching the potential of the MSB capacitors during the sampling phase. Theoretically, it enables a precision improvement of more than 0.5-bit. In this design, 512 capacitor units and a reference voltage $V_{REF}$ are employed, achieving an extended input range of $\pm 1.5 V_{REF}$ and an equivalent resolution of 10.5-bit. Fabricated in a 180-nm CMOS process, the prototype chip demonstrates 10.36-bit ENOB in post-simulation, while consuming 48μW at a sampling rate of 2 MS/s under a 1.8-V supply, with an area of 0.79mm².
**key words:** successive approximation register analog-to-digital converter (SAR ADC), precision-improvement technology, top-plate sampling, $V_{CM}$-based switching scheme
**Classification:** Integrated circuits

## 1. Introduction

In modern CMOS technologies, successive approximation register (SAR) analog-to-digital converters (ADCs) have attracted significant attention due to their excellent power efficiency, particularly achieving state-of-the-art performance in medium-resolution applications [1-3]. However, in many practical applications such as sensor readout, biomedical signal processing, and digital instrumentation, the demand for higher-resolution ADCs continues to grow.

The capacitive digital-to-analog converter (CDAC), which incurs no static power consumption and is highly compatible with advanced CMOS processes, has become the core building block in SAR ADCs. [4-6] The CDAC plays a decisive role in determining the power efficiency of the ADC. On the one hand, the number of unit capacitors directly defines the effective resolution of the ADC. On the other hand, the CDAC inherently determines the noise level of the SAR ADC. [7-9] From a resolution perspective, increasing the effective number of bits (ENOB) inevitably requires a larger CDAC. However, this increase comes at the expense of higher power consumption and larger silicon area, creating a fundamental design trade-off that cannot be easily avoided.

To further enhance the power efficiency of SAR ADCs and to decouple their resolution from the direct dependence on CDAC area, numerous techniques have been proposed, such as LSB repeat, oversampling, and $kT/C$ noise reduction. [10-15] While effective to some extent, each of these methods comes with inherent limitations. For instance, the LSB repeat technique improves resolution but requires additional conversion cycles, thereby restricting the maximum sampling rate. Oversampling enhances precision by averaging multiple digital outputs. However, it breaks the one-to-one mapping between input and output samples, which is undesirable in many applications. The $kT/C$ noise reduction techniques heavily depend on the performance of preamplifiers. It is worth noting that these approaches are particularly effective in high-resolution ADCs. In contrast, for medium-resolution SAR ADCs, where the architecture itself is the dominant choice due to its efficiency, the improvements from these techniques are often less significant.

Another approach to improving SAR ADC resolution is to extend its effective input range (*IR*). By allowing the input signal swing to exceed the conventional *IR*, the noise constraint on the circuit can be relaxed, thereby improving the signal-to-noise ratio (SNR). In [16], a dual-sub-ADC architecture was proposed, where each sub-ADC is responsible for processing half of the input range. It suffers from significant drawbacks, specifically, the need for two additional sets of hardware circuits, which substantially increase the chip area. In [17], an MSB prediction technique was introduced. In this design, the MSB capacitors are flipped based on a prediction, this effectively doubles the ADC's equivalent input range. However, since the MSB toggling relies on prediction, additional capacitors must be incorporated to determine the range of the input signal, which inevitably increases circuit complexity and area overhead.

[1] School of Advanced Manufacturing, Fuzhou University, Quanzhou 362251, China;
a) 852203429@fzu.edu.cn

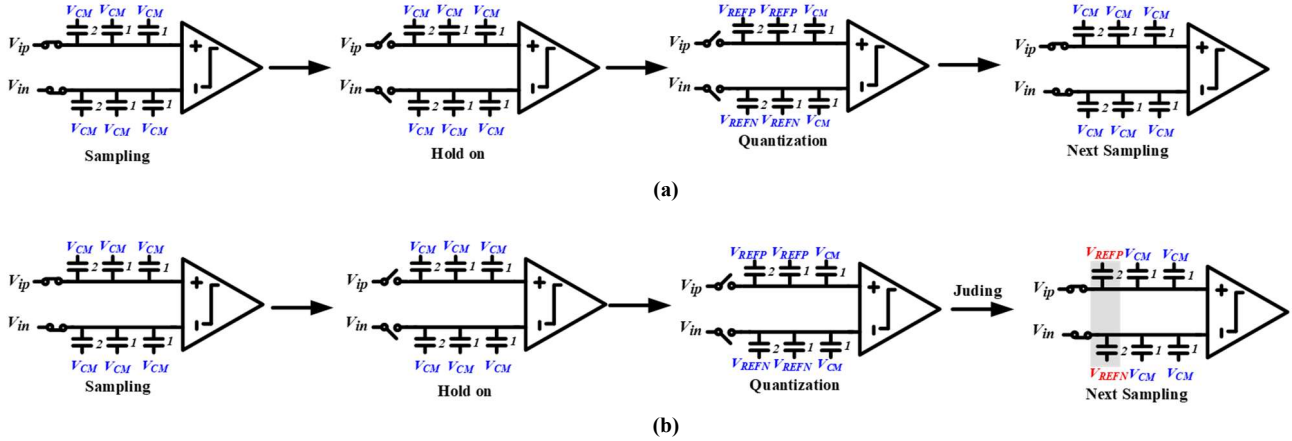

**Fig. 1.** (a). the conventional top-plate sampling scheme, (b) the top-plate sampling with extending input range.

Unlike the aforementioned technology, the scheme proposed in this work extends the input range only when the input exceeds 0.5×IR, by switching the MSB capacitor state during the sampling phase. This technique achieves a 1.5×IR extension without additional capacitors, thereby avoiding extra area and power overhead. With the enlarged input range, the effective resolution of the ADC is improved by over 0.5-bit, while the area and power consumption remains negligible. To validate the concept, a 10-bit SAR ADC with a sampling rate of 2 MS/s was designed. Post-layout simulation results demonstrate an effective resolution of 10.36-bit, with chip area of 0.79 mm².

This paper is organized as follows. Section II presents the principle of the proposed input range extension and the precision improvement technique. Section III describes the overall architecture of the ADC. Section IV details the control logic and critical circuit blocks. Section V discusses the post-layout simulation results of the prototype design.

## 2. Input Range Extension and Precision Improvement Technique

The technique proposed in this work improves the precision of SAR ADCs by expanding the effective input range. Under the same reference voltage $V_{REF}$, a larger effective input range allows the SAR ADC to achieve a higher signal-to-noise ratio (SNR), which can be expressed as:

$$\Delta SNR = SNR_{EXT} - SNR = 20\log\frac{V_{\sin\_ext}}{V_{\sin}} \quad (1)$$

$$\Delta ENOB = \log\frac{V_{\sin\_ext}}{V_{\sin}} \quad (2)$$

The proposed input range extension scheme with top-plate sampling and its switching logic are explained as an example in Fig. 1. In the conventional scheme, as shown in Fig. 1(a), the top-plate sampling switches sample the input signal while the bottom-plate switches are connected to $V_{CM}$. After the sampling phase, the top-plate switch is turned off, while the bottom-plate switches remain connected to $V_{CM}$. At this moment, the differential voltage sampled on the top plates corresponds to the input difference, $V_{ip} - V_{in}$.

In the proposed input range extension technique, the initial quantization result is used to determine whether the input signal exceeds $0.5 V_{REF}$. If this condition is met, the input range extension mechanism is activated in the next cycle. The decision-making process is carried out by monitoring the digital output code $D_{OUT}$ of the ADC. Specifically, when both MSB capacitors output logic "1," it indicates that the input signal amplitude exceeds $0.5 V_{REF}$. At this point, the ADC determines that input range extension is required. During the subsequent sampling phase, the positive-side MSB capacitor samples $V_{REFP}$, while the negative-side MSB capacitor samples $V_{REFN}$, thereby effectively reducing the equivalent input range.

This process decreases the sampled charge at the positive terminal and increases the sampled charge at the negative terminal, as expressed in (3) and (4).

$$Q_{S,P} = (V_{ip} - V_{REFP}) \times 2C + (V_{ip} - V_{CM}) \times 2C \quad (3)$$

$$Q_{S,N} = (V_{in} - V_{REFN}) \times 2C + (V_{in} - V_{CM}) \times 2C \quad (4)$$

The voltage difference at the top plates of the ADC comparator can be expressed as:

$$V_{CMP\_P} - V_{CMP\_N} = V_{ip} - V_{in} - 0.5 \times V_{REF} \quad (5)$$

That is, the input signal range of the ADC is extended to $1.5 V_{REF}$. According to (2), the effective resolution of the ADC is thereby improved by approximately 0.5-bit.

## 3. ADC Architecture

Since the proposed input range extension technique is primarily targeted at medium-resolution ADCs, a 10-bit $V_{CM}$-based differential SAR ADC prototype has been designed to validate its effectiveness. The prototype employs a top-plate sampling scheme and consists of sampling switches, a CDAC, and SAR control logic.

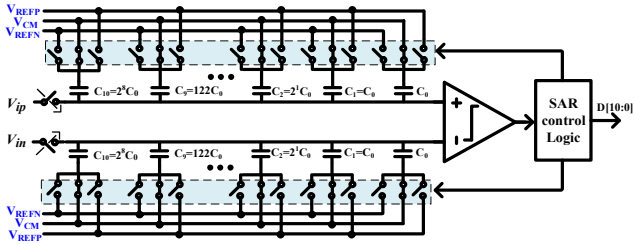

**Fig. 2.** ADC architecture

### 3.1 CDAC Redundancy Design
In SAR ADCs, the CDAC is typically implemented as a binary-weighted array. However, in practical measurements, capacitor mismatch is unavoidable. Since this work targets a medium-resolution ADC, calibration algorithms are rarely employed at this resolution level. To alleviate the impact of capacitor mismatch on ADC linearity, the proposed ADC adopts a redundancy-based design as expressed in (6). [18-19]

$$C_i < C_{i-1} + C_{i-2} + ... + C_1 + C_0 \qquad i \le N \qquad (6)$$

The principle of redundancy is to introduce an overlap in the search range during the conversion process. When a comparison error occurs, as long as the error falls within the overlapping range, it can be corrected by subsequent conversions, thereby ensuring that the final digital output remains accurate. Consequently, an $N$-bit ADC requires more than $N$ conversion steps to complete quantization. The capacitor array design is summarized in Table I, where capacitor $C_5$ serves as the redundancy bit to prevent errors caused by incorrect decisions in the higher capacitors.

### 3.2 Extension design

For a differential SAR ADC with a reference voltage of $V_{REF}$, the nominal quantization range is $[-V_{REF},+V_{REF}]$. In the proposed range-extension scheme, the initial quantization process is identical to that of a conventional top-plate sampled differential SAR ADC, as illustrated in Fig. 1(a). Once the two MSB output codes, D10 and D9, are both equal to "1", it indicates that the input signal amplitude exceeds $0.5V_{REF}$. In the subsequent sampling cycle, the bottom-plate switch of capacitor $C_{10}$ on the positive side is connected to $V_{REFP}$, while that on the negative side is connected to $V_{REFN}$, with all other capacitor bottom-plate switches held at $V_{CM}$. After sampling, all bottom-plate switches are then reset to $V_{CM}$. According to the derivation in Section II, the quantization range of the ADC is thus extended to $1.5V_{REF}$. Similarly, when both $D_{10}$ and $D_9$ are "00" the input amplitude is smaller than $-0.5V_{REF}$. In this case, during the next sampling cycle, the bottom-plate switch of capacitor $C_{10}$ on the positive side is connected to $V_{REFN}$, while that on the negative side is connected to $V_{REFP}$. This operation extends the input range to $-1.5V_{REF}$. Consequently, for a differential ADC, the effective quantization range is expanded to $[-1.5V_{REF},+1.5V_{REF}]$.

**Table I** The redundancy range for each bit.

| Cap number | weight | Redundant range/LSB |
|---|---|---|
| C10 | 256 | 0 |
| C9 | 122 | 8 |
| C8 | 62 | 10 |
| C7 | 30 | 12 |
| C6 | 16 | 12 |
| C5(redundancy) | 11 | 1 |
| C4 | 7 | 2 |
| C3 | 4 | 0 |
| C2 | 2 | 0 |
| C1 | 1 | 0 |
| C0 | 1 | 0 |

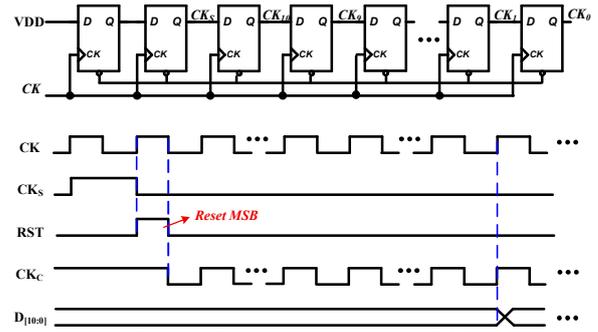

**Fig. 3.** Timing of the proposed ADC.

By keeping $V_{REF}$ and the quantization step constant, the proposed MSB capacitor switching during the sampling phase extends the effective input range from $\pm V_{REF}$ to $\pm 1.5V_{REF}$, the improvement of ENOB is more than 0.5-bit. In this way, it can be approximately regarded as a 0.5-bit enhancement in practical implementation, achieved without requiring additional capacitors or incurring significant power overhead.

### 4. Circuit Design and Implementation

#### 4.1 Timing
As shown in Fig. 3, the ADC control logic is implemented with fully synchronous digital circuits to simplify the design. Eleven clock pulses, $CK_1$ to $CK_{11}$, drive eleven *DFFs* to capture the outputs of the comparator. At the falling edge of $CK_C$, the dynamic comparator evaluates the input signal, and at the subsequent rising edge, only one *DFF* latches the comparator output. Thus, half of the clock cycle is allocated for the comparator decision, while the other half is used for CDAC voltage adjustment and comparator reset. The timing in this design differs from conventional top-plate sampling. In conventional top-plate sampling, the first comparison can be performed immediately after sampling. However, in this work, since the MSB capacitor is repurposed for input-range extension, additional time is required to reset the MSB capacitor after the sampling phase. Although this introduces a clock cycle overhead, the resulting input-range extension and effective resolution improvement make the trade-off worthwhile.

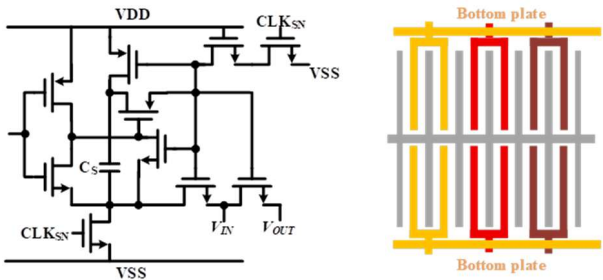

**Fig. 4.** (left) the bootstrap switch, (right) the MOM capacitor

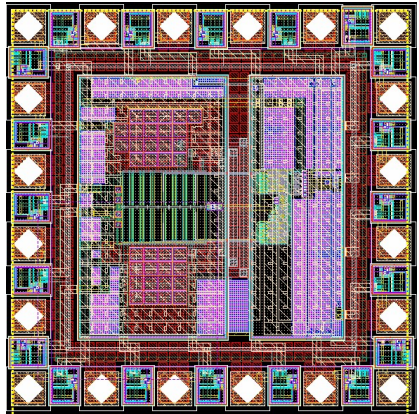

**Fig. 5.** Chip layout picture.

### 4.2 Other Circuit Design
The remaining circuit blocks follow conventional implementations [20-21]. To ensure high-performance sampling, bootstrapped switches are employed in the ADC to improve linearity. Considering the resolution requirement, a unit capacitance of 2fF is selected, implemented with a custom-designed MOM capacitor. The capacitor array shares a common top-plate, while each unit capacitor is surrounded by its bottom-plate to suppress parasitic effects. Centroid matching is further applied in the CDAC layout to mitigate capacitor mismatch, thereby enhancing overall ADC linearity.

### 5. Post-simulation results

The proposed SAR ADC is implemented in a 180-nm CMOS process, occupying a chip area of 0.79 mm². The chip layout is shown in Fig. 5, where the CDAC array dominates the area, followed by the digital control circuits, while the comparator and switch circuits occupy relatively smaller portions. The ADC features a nominal full-scale input range of $\pm V_{REF}$ and supports two

**Table II** Power Dissipation and Breakdown at 2MS/s

|  | 10-bit mode | 10.5-bit mode |
|---|---|---|
| Reference Power (μW) | 8.6 | 9.1 |
| $V_{CM}$ Power (μW) | 1.6 | 1.6 |
| Analog Power (μW) | 10.6 | 10.4 |
| Digital Power (μW) | 26.4 | 27.3 |
| Total Power (μW) | 47.2 | 48.4 |

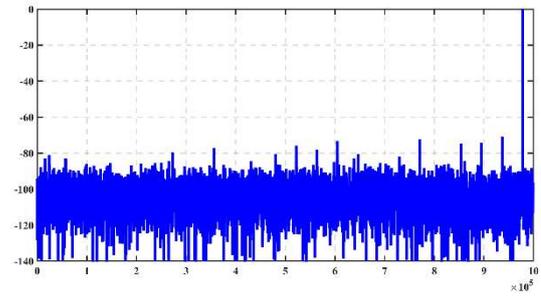

**(a)**

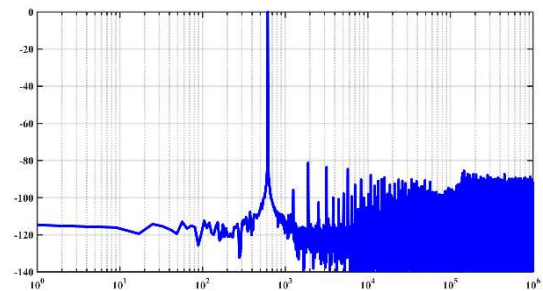

**(b)**

**Fig. 6.** The FFT spectrum of the (a)10-bit mode and (b)10.5-bit mode

operation modes: 10-bit mode with $\pm V_{REF}$ input range and 10.5-bit extended mode with $\pm 1.5 V_{REF}$ input range. In the standard $\pm V_{REF}$ mode, with a 999kHz sinusoidal input, the ADC achieves an SNDR of 60dB and an SFDR of 71.2dB. In the extended $\pm 1.5 V_{REF}$ mode, with a 10kHz sinusoidal input, the ADC achieves an SNDR of 64.1dB and an SFDR of 71dB. These dynamic test results verify the effectiveness of the proposed precision enhancement technique. A summary of the power consumption parameters is presented in Table II.

### 7. Conclusion

In this work, a 10-bit SAR ADC with 1.5× input range has been proposed and implemented in a 180-nm CMOS process. By introducing an MSB-capacitor switching during the sampling phase, the ADC effectively extends its input range from $\pm V_{REF}$ to $\pm 1.5 V_{REF}$, achieving an equivalent resolution improvement of approximately 0.5-bit without incurring significant area and power overhead. Post-simulation results validate the effectiveness of the proposed method, showing an improvement in dynamic performance when operating in the extended mode. The fabricated prototype achieves an SNDR of 64.1dB and an SFDR of 71dB in the extended mode, while consuming only a small chip area of 0.79mm². These results demonstrate that the proposed technique provides a simple yet efficient means of enhancing SAR ADC precision, making it well-suited for medium-resolution, low-power applications where area and efficiency are critical.